# Control of magnetization dynamics by substrate orientation in YIG thin films

Ganesh Gurjar[1], Vinay Sharma[3], S. Patnaik[1*], Bijoy K. Kuanr[2,*]

[1]School of Physical Sciences, Jawaharlal Nehru University, New Delhi, INDIA 110067

[2]Special Centre for Nanosciences, Jawaharlal Nehru University, New Delhi, INDIA 110067

[3]Morgan State University, Department of Physics, Baltimore, MD, USA 21251

## Abstract

Yttrium Iron Garnet (YIG) and bismuth (Bi) substituted YIG ($Bi_{0.1}Y_{2.9}Fe_5O_{12}$, BYG) films are grown in-situ on single crystalline Gadolinium Gallium Garnet (GGG) substrates [with (100) and (111) orientations] using pulsed laser deposition (PLD) technique. As the orientation of the Bi-YIG film changes from (100) to (111), the lattice constant is enhanced from 12.384 Å to 12.401 Å due to orientation dependent distribution of $Bi^{3+}$ ions at dodecahedral sites in the lattice cell. Atomic force microscopy (AFM) images show smooth film surfaces with roughness 0.308 nm in Bi-YIG (111). The change in substrate orientation leads to the modification of Gilbert damping which, in turn, gives rise to the enhancement of ferromagnetic resonance (FMR) line width. The best values of Gilbert damping are found to be $(0.54\pm0.06)\times10^{-4}$, for YIG (100) and $(6.27\pm0.33)\times10^{-4}$, for Bi-YIG (111) oriented films. Angle variation ($\phi$) measurements of the $H_r$ are also performed, that shows a four-fold symmetry for the resonance field in the (100) grown film. In addition, the value of effective magnetization ($4\pi M_{eff}$) and extrinsic linewidth ($\Delta H_0$) are observed to be dependent on substrate orientation. Hence PLD growth can assist single-crystalline YIG and BYG films with a perfect interface that can be used for spintronics and related device applications.

# 1. Introduction

One of the most widely studied materials for the realization of spintronic devices appears to be the iron garnets, particularly the yttrium iron garnet (YIG, $Y_3Fe_5O_{12}$) [1,2]. In thin film form of YIG several potential applications have been envisaged that include spin-caloritronics [3,4], magneto-optical (MO) devices, and microwave resonators, circulators, and filters [5–8]. The attraction of YIG over other ferroic materials is primarily due to their strong magneto-crystalline anisotropy and low magnetization damping [2]. Furthermore, towards high frequency applications, YIG's main advantages are its electrically insulating behavior along with low ferromagnetic resonance line-width (ΔH) and low Gilbert damping parameter [9–11]. These are important parameters for potential use in high frequency filters and actuators [12–14]. In this paper, we report optimal growth parameters for pure and Bi-doped YIG on oriented substrates and identify the conditions suitable for their prospective applications.

In literature, YIG is known to be a room temperature ferrimagnetic insulator with a $T_c$ near 560 K [15]. It has a cubic structure (space group $Ia\bar{3}d$). The yttrium (Y) ions occupy the dodecahedral 24c sites (in the Wyckoff notation), two Fe ions at octahedral 16a and three at tetrahedral 24d sites, and oxygen the 96h sites [16,17]. The d site is responsible for the ferrimagnetic nature of YIG. It is already reported that substitution of Bi in place of Y in YIG leads to substantial improvement in the magneto-optical response[7,18–25]. It was also observed that MO performance increases linearly with Bi/Ce doping concentration [22]. Furthermore, substitution of Bi in YIG (BYG) is documented to provide growth-induced anisotropy that is useful in applications such as magnetic memory and logic devices [26–30]. The study of basic properties of Bi-substituted YIG materials is of great current interest due to their applications in magneto-optical devices, magnon-spintronics, and related fields such as caloritronics due to its high uniaxial anisotropy and faraday rotation [21,31–35]. The structural and magnetic properties can be changed via change in $Bi^{3+}$ concentration in YIG or via choosing a proper substrate orientation. Therefore, the choice of perfect substrate orientation is crucial for the identification of the growth of Bi substituted YIG thin films.

In this work, we have studied the structural and magnetic properties of Bi-substituted YIG [$Bi_{0.1}Y_{2.9}Fe_5O_{12}$ (BYG)] and YIG thin films with two different single crystalline Gadolinium Gallium Garnet (GGG) substrate orientations: (100) and (111). The YIG and BYG films of thickness ~150 nm were grown by pulsed laser deposition (PLD) method [23,36,37] on top of single-crystalline GGG substrates. The structural and magnetic properties of all grown films were carried out using x-ray diffraction (XRD), surface morphology by atomic force microscopy (AFM), and magnetic properties via vibrating sample magnetometer (VSM) and ferromagnetic resonance (FMR) techniques. The FMR is the most useful technique to study the magnetization dynamics by measuring the properties of magnetic materials through evaluation of their damping parameter and linewidth. Furthermore, it provides insightful information on the static magnetic properties such as the saturation magnetization and the anisotropy field. FMR is also extremely helpful to study fundamentals of spin wave dynamics and towards characterizing the relaxation time and Lande *g* factor of magnetic materials [11].

## 2. Experiment

YIG and BYG targets were synthesized via the solid-state reaction method. Briefly, yttrium oxide ($Y_2O_3$) and iron oxide ($Fe_2O_3$) powders were ground for ~14 hours before calcination at 1100 ºC.

The calcined powders were pressed into pellets and sintered at 1300 ºC. Using these YIG and BYG targets, thin films of thickness ~150 nm were grown in-situ on (100)- and (111)-oriented GGG substrates by the PLD technique. The prepared samples have been labeled as YIG (100), YIG (111), BYG (100), and BYG (111). GGG substrates were cleaned using acetone and isopropanol. Before deposition, the deposition chamber was thoroughly cleaned and evacuated to a base vacuum of $2\times10^{-6}$ mbar. We have used KrF excimer laser (248 nm), with pulse frequency 10 Hz to ablate the targets at 300mJ energy. During deposition, target to substrate distance, substrate temperature, and oxygen pressure were kept at ~4.8 cm, 825 ºC, and 0.15 mbar, respectively. Best films were grown at a rate of 6 nm/min. The as-grown thin films were annealed in-situ for 2 hours at 825 ºC and cooled down to 300 ºC in the presence of oxygen (0.15 mbar) throughout the process. The structural properties of the thin film were determined by XRD using Cu-K$_\alpha$ radiation (1.5406 Å) and surface morphology as well as the thickness of the film were calculated with atomic force microscopy by WITec GmbH, Germany. Magnetic properties were studied using a 14 tesla PPMS (*Cryogenic)*. FMR measurements were carried out by the Vector Network Analyzer (VNA) (Keysight, USA) using a coplanar waveguide (CPW) in a flip-chip geometry with dc magnetic field applied parallel to the film plane.

## 3. Results and Discussion

### 3.1 Structural properties

The room temperature XRD data for the polycrystalline targets of YIG and BYG are plotted in figure 1 (a) and 1 (b) respectively. Rietveld refinement patterns after fitting XRD data are also included in the panels. XRD peaks are indexed according to the JCPDS card no. (# 43-0507). Inset of figure 1 (a) shows crystallographic sub-lattices of YIG that elucidates $Fe_1^{3+}$ tetrahedral site, $Fe_2^{3+}$ octahedral site, and $Y^{3+}$ dodecahedral site. Inset (i) of figure 1 (b) shows evidence for successful incorporation of Bi into YIG; the lattice constant increases when Bi is substituted into YIG due to larger ionic radii of Bi (1.170 Å) as compared with Y (1.019 Å) [19]. From Rietveld refinement we estimate the lattice constant of YIG and BYG to be 12.377 Å [38] and 12.401 Å respectively.

Figure 2 (a) and 2 (b) show the XRD pattern of bare (100) and (111) oriented GGG substrates. This is followed by figure 2 (c) & 2(d) for YIG and figure 2 (e) & 2 (f) for BYG as grown thin films. XRD patterns confirm the single-crystalline growth of YIG and BYG thin films over GGG substrates. The lattice constant, lattice mismatch (with respect to substrate), and lattice volume obtained from XRD data are listed in Table 1. Lattice constant *a* for the cubic structure is evaluated using the [39].

$$a = \frac{\lambda\sqrt{h^2+k^2+l^2}}{2\sin\theta} \quad (1)$$

Where $\lambda$ is the wavelength of Cu-K$\alpha$ radiation, $\theta$ is the diffraction angle, and [h, k, l] are the miller indices of the corresponding XRD peak. Lattice misfit ($\frac{\Delta a}{a}$) is evaluated using equation 2 [24,38].

$$\frac{\Delta a}{a} = \frac{(a_{film} - a_{substrate})}{a_{film}} \times 100\% \quad (2)$$

Where $a_{film}$ and $a_{substrate}$ are the lattice constant of film and substrate respectively. Lattice constant of pure YIG bulk is 12.377 Å, whereas we have observed a larger value of lattice constants of YIG and BYG films than that of bulk YIG as shown in Table 1. Similarly, to these obtained results, a larger value of lattice constants than that of bulk YIG has been reported as well [40–44].

The obtained values of the lattice constant are in agreement with the previous reports [18,21,25,34,45]. In the case of BYG (111), the value of lattice constant slightly increases compared to BYG (100) because the distribution of $Bi^{3+}$ in the dodecahedral site depends on the orientation of the substrate [28,46]. Inset (ii) of figure 1 (b) shows plane (111) has more contribution of $Bi^{3+}$ ions [(ionic radius of bismuth (1.170 Å) is larger as compared with YIG (1.019 Å) [19]]. This slight increase in the lattice constant (in 111 direction) implies a more lattice mismatch (or strain) in BYG films. Positive value of lattice mismatch indicates the slightly larger lattice constant of films (YIG and BYG) were observed as compared to substrates (GGG). We would like to emphasize that lattice plane dependence growth is important to signify the changes in the structural and magnetic properties.

### 3.2 Surface morphology study

Room temperature AFM images with roughness are shown in figure 3 (a)-(d). Roughness plays an important role from the application prospective as it is related to Gilbert damping factor α. Lower roughness (root mean square height) is observed for the (111) oriented films of YIG and BYG compared to those grown on (100) oriented substrates. Available literature [61] indicate that roughness would depend more on variation on growth parameters rather than on substrate orientation.  In this sense further study is needed to clarify substrate dependence of roughness. No significant change in the roughness is observed between YIG and BYG films [38,47]. Table 1 depicts a comparison between the roughness measured in YIG and the BYG thin films.

### 3.3 Static magnetization properties

VSM magnetization measurements were performed at 300 K with magnetic field applied parallel to the film plane (in-plane). Figure 3 (e) and 3 (f) show the magnetization plots for YIG and BYG respectively after careful subtraction of paramagnetic contribution that is assigned to the substrate. The measured saturation magnetization ($4\pi M_S$) data are given in Table 1 which are in general agreement with the reported values [11,40,48]. Not much change in the measured $4\pi M_S$ value of YIG and Bi-YIG films are observed. The ferrimagnetism nature of YIG arises from super-exchange interaction between the non-equivalent $Fe^{3+}$ ions at octahedral and tetrahedral sites [49]. Bismuth located at dodecahedral site does not affect the tetrahedral and octahedral $Fe^{3+}$ ions. So, Bismuth does not show a significant change in saturation magnetization at room temperature. It is reported in literature that Bi addition leads to increase in Curie temperature, so in that sense there is an decreasing trend in saturation magnetization in BYG films in contrast to YIG films [50,51]. Error bars in saturation magnetization relate to uncertainty in sample volume.

### 3.4 Ferromagnetic resonance properties

The FMR absorption spectroscopy is shown in figure 4. These measurements were performed at room temperature. The external dc magnetic field was applied parallel to the plane of the film. Lorentzian fit of the calibrated experimental data are used to calculate the FMR linewidth ($\Delta H$) and resonance magnetic field ($H_r$). From the ensemble of all the FMR data at different resonance frequencies (f = 1 GHz-12 GHz), we have calculated the gyromagnetic ratio ($\gamma$), effective magnetization field ($4\pi M_{eff}$) from the fitting of Kittel's in-plane equation [52].

In general, the uniform precession of magnetization can be described by the Landau-Lifshitz-Gilbert (LLG) equation of motion;

$$\frac{\partial \vec{M}}{\partial t} = -\gamma(\vec{M} \times \vec{H}_{eff}) + \frac{G}{\gamma M_s^2}\left[\vec{M} \times \frac{\partial \vec{M}}{\partial t}\right] \qquad (3)$$

Here, the first term corresponds to the precessional torque in the effective magnetic field and the second term is the Gilbert damping torque. The gyromagnetic ratio is given by $\gamma = g\mu_B/\hbar$, where $g$ is the Lande's factor, $\mu_B$ is Bohr magnetron and $\hbar$ is the Planck's constant. Similarly, $G = \gamma \alpha M_s$ is related to the intrinsic relaxation rate in the nanocomposites and $\alpha$ represents the Gilbert damping constant. $M_s$ (or $4\pi M_s$) is the saturation magnetization. It can be shown that the solution for in-plane resonance frequency can be written as;

$$f_r = \gamma'\sqrt{(H_r)(H_r + 4\pi M_{eff})} \qquad (4),$$

Where $\gamma' = \gamma/2\pi$, $4\pi M_{eff} = 4\pi M_s - H_{ani}$ is the effective field and $H_{ani} = \frac{2K_1}{M_s}$ is the anisotropy field. Following through, we have obtained Gilbert damping parameter (α) and inhomogeneous broadening (ΔH₀) linewidth from the fitting of Landau–Lifshitz–Gilbert equation (LLG) [53]

$$\Delta H(f) = \Delta H_0 + \frac{4\pi\alpha}{\sqrt{3}\gamma}f \qquad (5)$$

Derived parameters from the FMR study are listed in Table 2. The obtained Gilbert damping (α) is in agreement with the reported thin films used for the study of spin-wave propagation [2,27,54,55]. In the case of YIG not much change in the value of α is seen. However, a substantial increase is observed in case of BYG with (111) orientation. Qualitatively this could be assigned to the presence of $Bi^{3+}$ ions which induces spin-orbit coupling (SOC) [56–58] and also due to electron scattering inside the lattice as lattice mismatch (or strain) increases [59]. We have seen more distribution of $Bi^{3+}$ ions along (111) planes (see inset (ii) of figure 1 (b)) and also slightly larger lattice mismatch in BYG (111) from our XRD results. These results also explain higher value of Gilbert damping and $\Delta H_0$ in case of BYG (111). The change in $4\pi M_{eff}$ could be attributed to uniaxial in-plane magnetic anisotropy. This is because no change in $4\pi M_S$ is observed from magnetization measurements and $4\pi M_{eff} = 4\pi M_s - H_{ani}$ [38,40,60]. The uniaxial inplane magnetic anisotropy is induced due to lattice mismatch between films and GGG substrates [38,40]. The calculated gyromagnetic ratio (γ) and $\Delta H_0$ are also included in Table 2. The magnitude of $\Delta H_0$ is close to reported values for same substrate orientation [38]. In summary we find that YIG with (100) orientation yields lowest damping factor and extrinsic contribution to linewidth. These are the required optimal parameters for spintronics application with high spin diffusion length. However, MOKE signal is usually very low in bare YIG thin films because of its lower magnetic anisotropy and strain [61]. But previous reports suggest that magnetic anisotropy and magnetic domains formation can be achieved in YIG system by doping rare earth materials like Bi and Ce [18,61]. We have shown that anisotropic characteristic with Bi doping in YIG is more pronounced along <111> direction which can lead to the enhanced MOKE signal in Bi-YIG films on <111> substrate.

We have also recorded polar angle (θ) data of resonance field ($H_r$) versus magnetic field (H) at frequency 12 GHz for the BYG (100) and BYG (111) films (figure 5 (c) & 5(d) respectively where inset shows the azimuthal angle (ϕ) variation of $H_r$ measured at frequency of 3 GHz). The data are fitted with modified Kittel equation. From figure 5 (c) & (d), we can see that $H_r$ increases up to 2.5 kOe in BYG (100) and 3.0 kOe in BYG (111) by varying the direction of H from 0 to 90 degree with respect to sample surface (inset of Fig 5 (a)). Obtained parameters from angular variation of FMR magnetic field $H_r(\theta_H)$ are listed in the inset of figure 5 (c) & (d). From ϕ variation data (by varying the direction of H from 0 to 180 degree with respect to sample edge (Fig 5 (a)) Inset), we see clear four-fold and two-fold in-plane anisotropy in BYG (100) and BYG (111) films [61,62]. This further consolidates single-crystalline characteristics of our films. The change observed in $H_r$ with respect to ϕ variation is 79.52 Oe in BYG (100) ($\phi_H$=0 to 45) and 19.25 Oe in BYG (111) ($\phi_H$=0 to 45). Thus, during in-plane rotation, higher change in FMR field is observed along (100) orientation.

## 4. Conclusion

In conclusion, we have grown high quality YIG and Bi-YIG thin films on GGG substrates with (100) and (111) orientation. The films were grown by pulsed laser deposition. The optimal parameters i.e. target to substrate distance, substrate temperature, and oxygen pressure are determined to be ~ 4.8 cm, 825 °C, and 0.15 mbar, respectively. The as grown thin films have smooth surfaces and are found to be phase pure from AFM and XRD characterizations. From FMR measurements, we have found lower value of damping parameter in (100) YIG that indicates higher spin diffusion length for potential spintronics application. On the other-hand bismuth incorporation to YIG leads to dominance of anisotropic characteristics that augers well for application in magnetic bubble memory and magneto-optic devices. The enhanced value of α in Bi-YIG films is ascribed to the spin orbit coupled $Bi^{3+}$ ions. We also tabulate the values of magnetic parameters such as linewidth ($\Delta H_0$), gyromagnetic ratio (γ), and effective magnetization $4\pi M_{eff}$ with respect to substrate orientation. Unambiguous four-fold in-plane anisotropy is observed in (100) oriented films. We find high-quality magnetization dynamics and lower Gilbert damping parameter is possible in Bi-YIG grown on (111) GGG in conjunction with enhanced magnetic anisotropy. The choice of perfect substrate orientation is therefore found to be crucial for the growth of YIG and Bi-YIG thin films for high frequency applications.

## Acknowledgments

This work is supported by the MHRD-IMPRINT grant, DST (SERB, AMT, and PURSE-II) grant of Govt. of India. Ganesh Gurjar acknowledges CSIR, New Delhi for financial support. We acknowledge AIRF, JNU for access of PPMS facility.

A perspective on synthesis methods, J. Magn. Magn. Mater. 439 (2017) 277–286.

[6] S.-Y.S.Y. Huang, X. Fan, D. Qu, Y.P.P. Chen, W.G.G. Wang, J. Wu, T.Y.Y. Chen, J.Q.Q. Xiao, C.L.L. Chien, Transport magnetic proximity effects in platinum, Phys. Rev. Lett. 109 (2012) 107204. doi:10.1103/PhysRevLett.109.107204.

[7] A. Sposito, S.A. Gregory, P.A.J. de Groot, R.W. Eason, Combinatorial pulsed laser deposition of doped yttrium iron garnet films on yttrium aluminium garnet, J. Appl. Phys. 115 (2014) 53102.

[8] J.C. Butler, J.J. Kramer, R.D. Esman, A.E. Craig, J.N. Lee, T. Ryuo, Microwave and magneto-optic properties of bismuth-substituted yttrium iron garnet thin films, J. Appl. Phys. 67 (1990) 4938–4940.

[9] A.A. Serga, A. V Chumak, B. Hillebrands, YIG magnonics, J. Phys. D. Appl. Phys. 43 (2010) 264002.

[10] W.T. Ruane, S.P. White, J.T. Brangham, K.Y. Meng, D. V Pelekhov, F.Y. Yang, P.C. Hammel, Controlling and patterning the effective magnetization in Y3Fe5O12 thin films using ion irradiation, AIP Adv. 8 (2018) 56007.

[11] T. Liu, H. Chang, V. Vlaminck, Y. Sun, M. Kabatek, A. Hoffmann, L. Deng, M. Wu, Ferromagnetic resonance of sputtered yttrium iron garnet nanometer films, J. Appl. Phys. 115 (2014) 87–90. doi:10.1063/1.4852135.

[12] S. Dai, S.A. Bhave, R. Wang, Octave-Tunable Magnetostatic Wave YIG Resonators on a Chip, IEEE Trans. Ultrason. Ferroelectr. Freq. Control. 67 (2020) 2454–2460.

[13] C.S. Tsai, G. Qiu, H. Gao, L.W. Yang, G.P. Li, S.A. Nikitov, Y. Gulyaev, Tunable wideband microwave band-stop and band-pass filters using YIG/GGG-GaAs layer structures, IEEE Trans. Magn. 41 (2005) 3568–3570.
12

**List of Tables with caption**

**Table 1:** Lattice and magnetic parameters obtained from XRD, AFM and VSM.

| S. No. | Sample | Lattice constant (Å) | Lattice Mismatch (%) | Lattice volume (Å$^3$) | Roughness (nm) | $4\pi M_S$ (Gauss) |
|---|---|---|---|---|---|---|
| 1. | YIG (100) | 12.403 | 0.42 | 1907.81 | 0.801 | 1670.15±83.51 |
| 2. | YIG (111) | 12.405 | 0.40 | 1909.02 | 0.341 | 1654.06±82.70 |
| 3. | BYG (100) | 12.384 | 0.36 | 1899.11 | 0.787 | 1788.50±89.43 |
| 4. | BYG (111) | 12.401 | 0.65 | 1906.93 | 0.308 | 1816.31±90.82 |

**Table 2:** Damping and linewidth parameters obtained from FMR

| S. No. | Sample | $\alpha$ (×10$^{-4}$) | $\Delta H_0$ (Oe) | $4\pi M_{eff}$ (Oe) | $\gamma'$ (GHz/kOe) |
|---|---|---|---|---|---|
| 1. | YIG (100) | (0.54±0.06) | 26.24±0.10 | 1938.60±37.57 | 2.89±0.01 |
| 2. | YIG (111) | (1.05±0.13) | 26.51±0.21 | 2331.38±65.78 | 2.86±0.02 |
| 3. | BYG (100) | (1.66±0.10) | 26.52±0.17 | 1701.67±31.87 | 2.89±0.11 |
| 4. | BYG (111) | (6.27±0.33) | 29.28±0.62 | 2366.85±62.60 | 2.86±0.02 |

**Figure Captions**

**Figure 1:** XRD with Rietveld refinement pattern of (a) YIG target (inset shows crystallographic sub-lattices, $Fe_1^{3+}$ tetrahedral site, $Fe_2^{3+}$ octahedral site and $Y^{3+}$ dodecahedral site) (b) BYG target (inset (i) shows effect of Bi doping into YIG, inset (ii) shows contribution of the $Bi^{3+}$ along the (100) and (111) planes).

**Figure 2:** XRD pattern of (a) GGG (100), (b) GGG (111), (c) YIG (100), (d) YIG (111), (e) BYG (100), and (f) BYG (111).

**Figure 3:** AFM images of (a) YIG (100), (b) YIG (111), (c) BYG (100), (d) BYG (111) and static magnetization graph of (e) YIG (100), YIG (111); and (f) BYG (100), BYG (111).

**Figure 4:** FMR absorption spectra of (a) YIG (100), (b) YIG (111), (c) BYG (100), and (d) BYG (111).

**Figure 5:** (a) FMR magnetic field $H_r$ is plotted as a function of frequency f. Experiment data fitted with Kittel equation for YIG and BYG oriented films. Inset shows how the applied field angle is measured from sample surface (b) Frequency-dependent FMR linewidth data fitted with LLG equation for YIG and BYG oriented films. Inset shows the magnified version to illustrate the effect of Bi doping in YIG. (c) and (d) show angular variation of FMR magnetic field ($H_r(\theta_H)$) fitted with modified Kittel equation at 12 GHz frequency for BYG (100) and BYG (111) films. Insets show the FMR magnetic field ($H_r$) as a function of azimuthal angle ($\phi$).

**Figure 1**

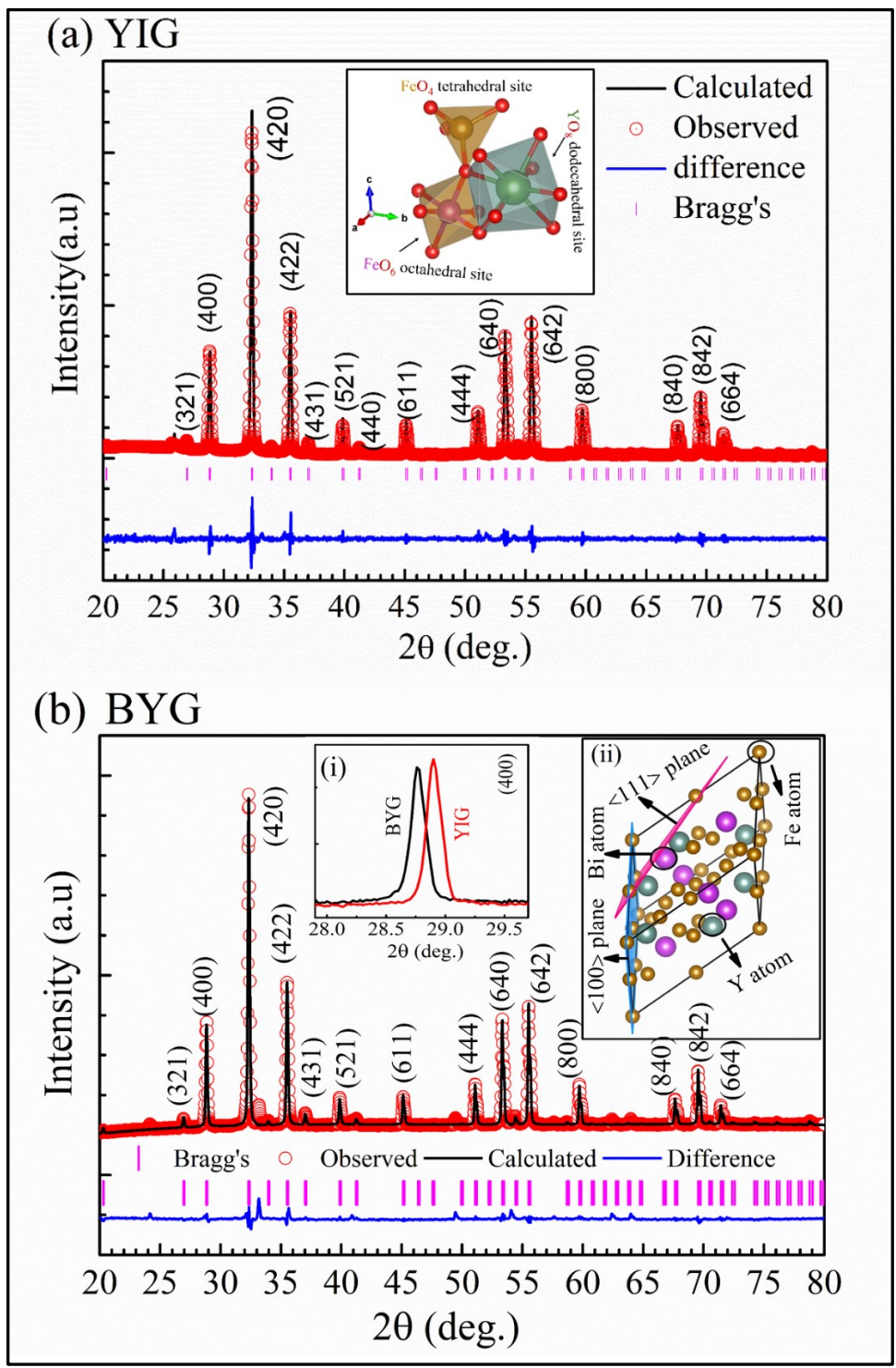

**Figure 2**

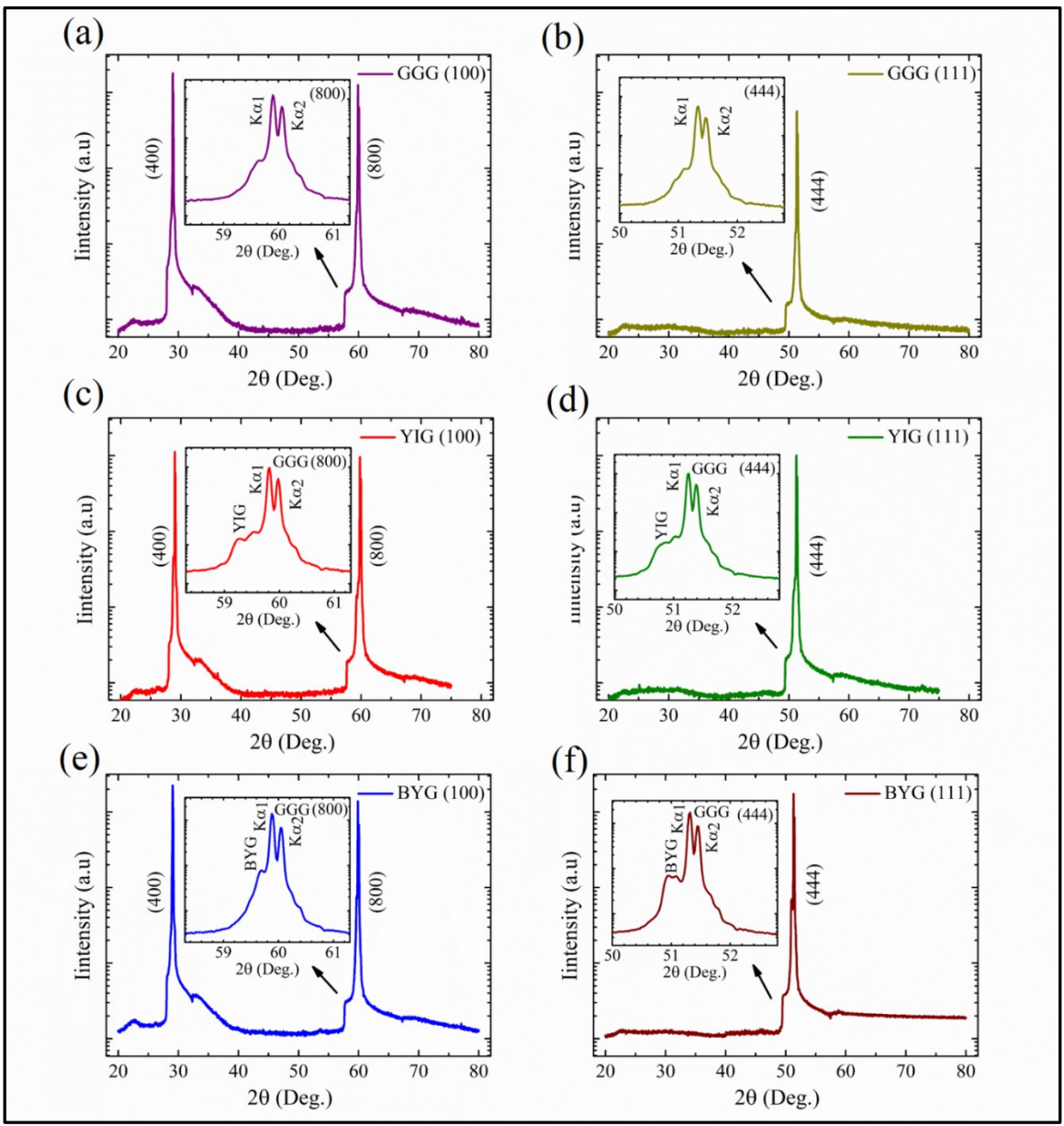

**Figure 3**

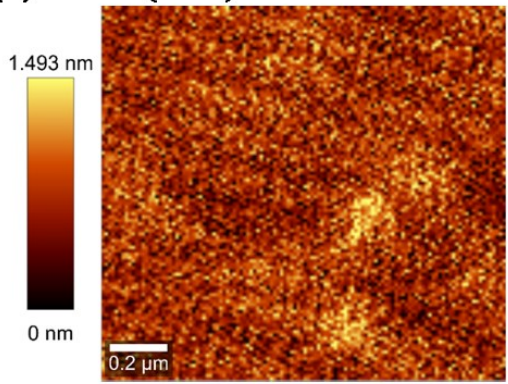
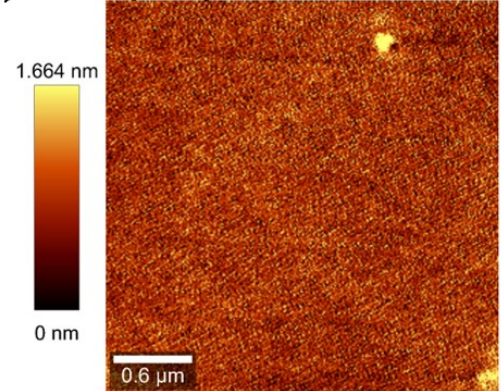
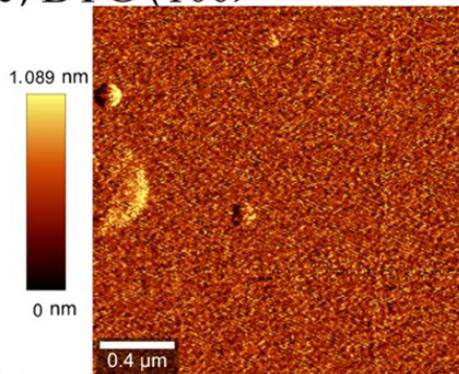
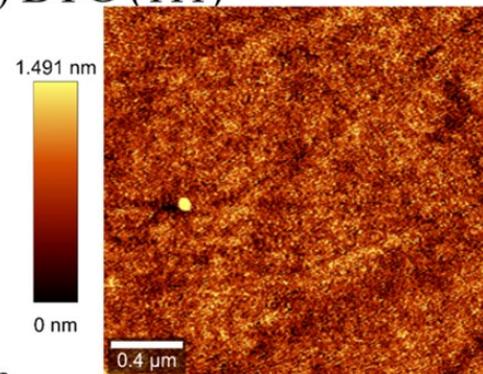
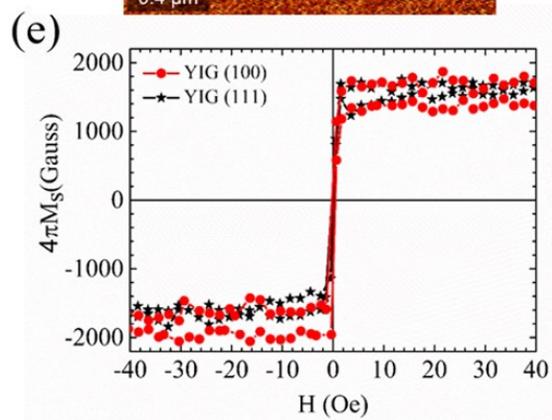
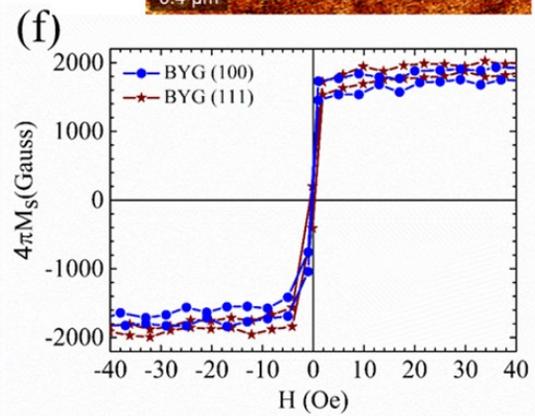

**Figure 4**

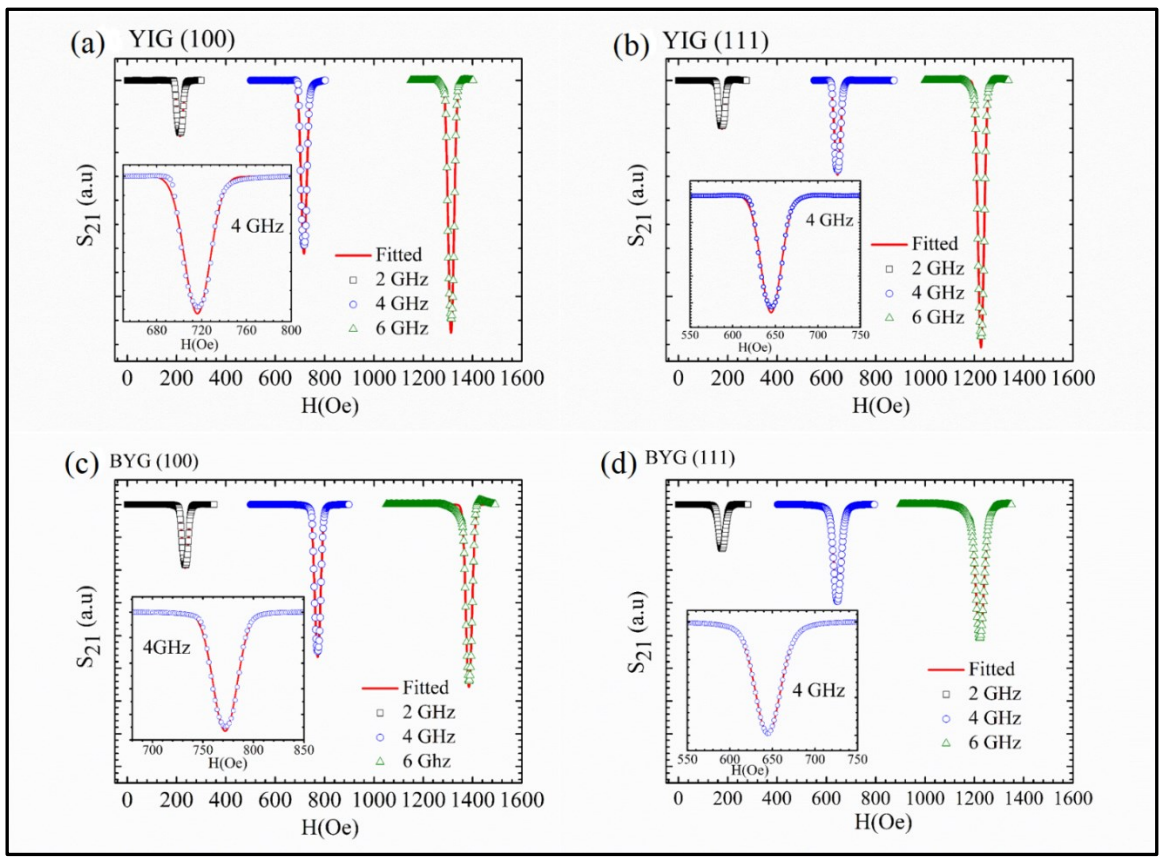

**Figure 5**

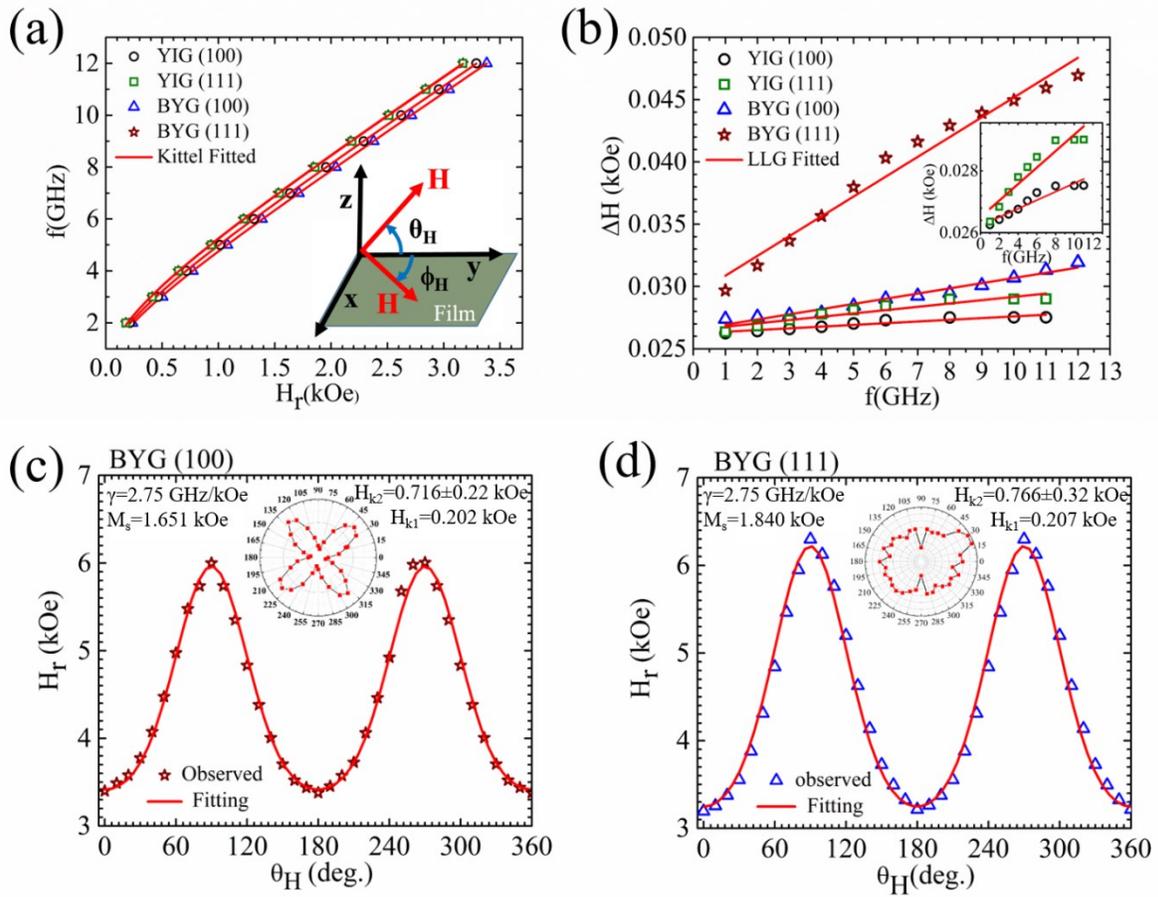